\documentclass[aps,prl,10pt,twocolumn,preprintnumbers,superscriptaddress,nofootinbib,floatfix]{revtex4-2}

\usepackage{graphicx,epstopdf}
\usepackage{amssymb,amsmath}
\usepackage{dcolumn}
\usepackage{bm}
\usepackage[dvipsnames]{xcolor}
\usepackage{color}
\usepackage{enumerate}
\usepackage[colorlinks=true,citecolor=blue]{hyperref}
\hypersetup{colorlinks=true,citecolor=blue,linkcolor=blue,urlcolor=blue}
\usepackage{soul}

\DeclareMathAlphabet\bmcal{OMS}{cmsy}{b}{n}

\usepackage{tabularx}
\usepackage{comment}

\begin{document}

\title{Acoustically-driven magnons in CrSBr bilayers}
\author{Anton Shubnic}
\affiliation{Department of Physics, ITMO University, Saint Petersburg 197101, Russia}
\author{Igor Chestnov}
\affiliation{Department of Physics, ITMO University, Saint Petersburg 197101, Russia}
\author{Igor Lobanov}
\affiliation{Department of Physics, ITMO University, Saint Petersburg 197101, Russia}
\author{Valery Uzdin}
\affiliation{Department of Physics, ITMO University, Saint Petersburg 197101, Russia}
\author{Ivan Iorsh}
\affiliation{Department of Physics, Engineering Physics and Astronomy, Queen's University, Kingston, Ontario K7L 3N6, Canada}
\affiliation{Department of Physics, ITMO University, Saint Petersburg 197101, Russia}
\author{Ivan A. Shelykh}
\affiliation{Science Institute, University of Iceland, Dunhagi 3, IS-107, Reykjavik, Iceland}

\date{\today}

\begin{abstract}
   We study the coupling between spin excitations and acoustic waves in bilayers of CrSBr, an ambiently stable 2D magnetic material.  We demonstrate that a strong dependence of inter-layer exchange coupling on strain makes possible the resonant generation of magnons by an acoustic wave. It is shown that the parameters of the generation, in particular the resonant frequency, can be tuned by an external  magnetic field, which makes CrSBr a promising platform for spintronics applications.
\end{abstract}

\maketitle

\textit{Introduction.}
Since the discovery of graphene \cite{novoselov2004electric}, the physics of two-dimensional materials has evolved in a rapidly developing field with prospects for applications in nanoelectronics. Layered van-der-Waals materials are characterized by strong light-matter \cite{luo2024strong} and spin-orbit interactions \cite{tang2021spin}, which makes them prominent candidates for optoelectronic and spintronics applications. For a long time, the question of the existence of magnetic two-dimensional materials remained open until CrI$_3$ \cite{huang2017layer}, an Ising ferromagnet with a Curie temperature of $\approx$45~K, and possessing  a set of outstanding properties, such as giant tunneling magnetoresistance \cite{kim2018one}, magnetoelectric coupling \cite{jiang2018controlling}, magnetic second harmonic generation \cite{sun2019giant} and excitonic Rabi splitting \cite{Zhumagulov2023} was experimentally discovered.
Unfortunately, this material is unstable under ambient conditions as it is subject to fast oxidation and hydration processes, which seriously limit the range of its possible applications \cite{huang2020emergent}. 

Recently, an air stable 2D material, namely CrSBr was discovered. Possessing an orthorombic symmetry, this material is characterized by two non-equivalent crystallographic directions and demonstrates a strong anisotropy of its optical \cite{wilson2021interlayer}, electronic \cite{wu2022quasi} and magnetic \cite{boix2022probing} properties. In monolayer geometry it is a ferromagnet with Curie temperature of about 140 K \cite{lee2021magnetic}. In multilayer configuration, the intra-layer ferromagnetic (FM) order coexists with the antiferromagnetic (AFM) inter-layer coupling \cite{telford2020layered}, so that in the zero external magnetic field the net magnetization remains zero. 

One of the key features of 2D magnetic materials is the possibility of tuning of their magnetic properties by external strains. For example, applying strain to the MoN$_2$ monolayer, it is possible to achieve a phase transition accompanied by switching the magnetic order from FM to AFM \cite{wang2016strain}. 
Recent experimental \cite{cenker2022reversible,Bagan2024} and theoretical \cite{liu2024intralayer} studies show that CrSBr exhibits a similar strong response to mechanical deformation. In particular, the uniaxial strain applied along the crystallographic $a$-axis dramatically modifies the interlayer exchange interaction, driving a transition from AFM to FM order \cite{cenker2022reversible}. This strong magneto-elastic coupling opens new avenues for exploring novel magneto-mechanical effects in CrSBr.

In particular, magnons, the elementary excitations in the spin subsystem of a material relevant for spintronics applications \cite{chumak2015magnon}, were reported to be efficiently coupled to phonons  \cite{sadovnikov2018magnon}. 
In this article, we build a microscopic theory of this remarkable effect, and demonstrate the possibility of resonant excitation of GHz magnons by an external acoustic wave due to the time-periodic modulation of the interlayer exchange. We show that the coupling strength between magnons and phonons is proportional to the cross product of the Neel and net magnetization vectors. 
Consequently, an external out-of-plane magnetic field is required to activate this coupling.
We analyze how the generation frequency, amplitude, and resonance half-width depend on the external magnetic field, highlighting the potential of CrSBr bilayers for precisely controllable spintronic devices.

\begin{figure}[h!]
    \centering
    \includegraphics[width = \linewidth]{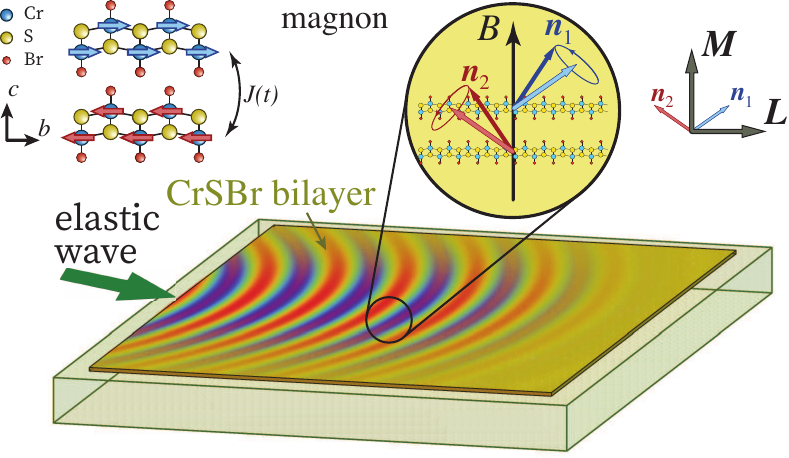}
    \caption{The geometry of the proposed device for resonant acoustic generation of magnons in CrSBr bilayer. An acoustic wave produces the in-plane strain in the CrSBr (color coded), which modulates the interlayer magnetic exchange parameter $J$. A periodic change in $J$ leads to the generation of a spin wave. The coupling strength between magnons and phonons is proportional to the cross product of the Neel and net magnetization vectors,  so the application of an external out-of-plane magnetic field is necessary.}
\label{bilayer}
\end{figure}

\textit{Model.}
We consider a bilayer of CrSBr as sketched in Fig.~\ref{bilayer}. The stable magnetic configuration of the system can be obtained by minimization of the following energy functional \cite{bae2022exciton}:
\begin{align}\label{eq:energy}
    & E_0 =  \int {\rm d}^2 {\bm r} \left[ \hat{D}[{\bm n_1}] + \hat{D}[{\bm n_2}] + \frac{K_c}{2} (n^2_{1c} + n^2_{2c}) + \right.
    \nonumber \\
    & \left. \frac{K_a}{2} (n^2_{1a} + n^2_{2a}) + J{\bm n_1 \bm n_2} - B(n_{1c} + n_{2c})  \right],
    \\
    & \hat{D}[{\bm n}] = A_a \left( \frac{\partial {\bm n}}{\partial a} \right)^2 + A_{ab} \frac{\partial {\bm n}}{\partial a} \frac{\partial {\bm n}}{\partial b} + A_b \left( \frac{\partial {\bm n}}{\partial b} \right)^2,
    \nonumber
\end{align}
where ${\bm n}_1$ and ${\bm n}_2$ are the magnetization unit vector of the upper and lower layers, respectively, $J\equiv J(t)$ is the interlayer exchange constant, sensitive to a local strain created by the elastic wave; $B$ is the external magnetic field applied along the hard axis $c$; $K_c$ and $K_a$ are single-ion anisotropies in $c$ and $a$ crystallographic directions. Here we use $K_c =0.9$~meV/nm$^2$, $K_a = 0.55$~meV/nm$^2$ and $J = 0.076$~meV/nm$^2$ \cite{bae2022exciton} in the equilibrium. Here we took into account that the effective interlayer magnetic exchange in the bilayer is twice as low as in a bulk. In addition, using the intralayer exchange coupling constants $J_a = 4.27$~meV, $J_b= 3.76$~meV, $J_{ab}= 7.6$~meV from \cite{scheie2022spin} we obtain the following values for the intralayer exchange parameters: $A_a = 13.36$~meV, $A_b =23.12$~meV, $A_{ab} = 17.1$~meV in the continuum limit (see Supplemental Material \cite{SM}). 

The spin dynamics is described by a set of Landau-Lifshitz-Gilbert (LLG) equations:
\begin{equation}\label{LLGeqs}
\left\{
\begin{aligned}
    \frac{\partial {\bm n_1}}{\partial t} &=  \frac{\gamma_0}{n_s}  \left[ {\bm n_1} \times \frac{ \delta E_0}{\delta {\bm n_1}} \right] + \alpha \left[ {\bm n_1} \times \frac{\partial {\bm n_1}}{\partial t} \right], \\
    \frac{\partial {\bm n_2}}{\partial t} &=  \frac{\gamma_0}{n_s} \left[ {\bm n_2} \times \frac{ \delta E_0}{\delta {\bm n_2}}  \right] + \alpha \left[ {\bm n_2} \times \frac{\partial {\bm n_2}}{\partial t} \right],
\end{aligned}
\right.
\end{equation}
where $\gamma_0 = 28$~GHz/T  and $\alpha = 0.066$ are correspondingly the gyromagnetic ratio and the Gilbert damping constant \cite{bae2022exciton,cham2025spin}; $n_s = (3/2) \mu_B/(a_0 b_0/2)$ is a saturation magnetization, where $\mu_B$ is the Bohr magneton,  $a_0= 0.351 $~nm and $b_0= 0.477$~nm are the lattice constants \cite{goser1990magnetic}, and we take into account the presence of two Cr atoms in a unit cell with three electrons in each atom.

Let us introduce the total magnetization ${\bm M} = {\bm n_1} + {\bm n_2}$ and the Neel  ${\bm L} = {\bm n_1} - {\bm n_2}$ vectors. At zero magnetic field, the bilayer is in the AFM state  $M=0$ and ${\bm L} = 2 {\bm{e}_b}$. The canted phase (see Fig.~\ref{bilayer}) with ${\bm{M}}=2B/\left(2J+K_c\right) {\bm{e}_c}$ and ${\bm{L}}=\sqrt{4-M^2}{\bm{e}_b}$ is realized at $B<B_c$, where $B_c = 2J + K_c$ is the critical field corresponding to the transition to the FM phase.

To determine the magnon dispersion in the CrSBr bilayer, we linearize the LLG equations \eqref{LLGeqs} by the spin-wave ansatz: ${\bm M} = {\bm M}_0 + {\bm m}e^{i(\omega t-{\bm q \bm r})}$, ${\bm L} = {\bm L}_0 + {\bm l}e^{i(\omega t-{\bm q \bm r})}$, and investigate the evolution of weak excitations ${\bm m}$ and ${\bm l}$ around the equilibrium values ${\bm M}_0$ and ${\bm L}_0$. Using the relations ${\bm M \bm L}=0$ and ${\bm M}^2 + {\bm L}^2 = 4$, which follows from the normalization conditions, reduces the problem to the decoupled evolution of the in-plane magnetization ${\bm m}_\perp = (m_a,m_b)$ and Neel vector ${\bm l}_\perp = (l_a,l_b)$ components:
\begin{subequations}
\begin{align}
&\left\{
\begin{aligned} \label{eq:mdyn}
    &\frac{dm_a}{dt} = - C^m_{ab} m_b - \Gamma^m_a m_a,
    \\
    &\frac{dm_b}{dt} = C^m_{ba} m_a - \Gamma^m_b m_b,
\end{aligned}
\right. \\
&\left\{
\begin{aligned}  \label{eq:ldyn}
    &\frac{dl_a}{dt} = - C^l_{ab} l_b - \Gamma^l_a l_a,
    \\
    &\frac{dl_b}{dt} = C^l_{ba} l_a - \Gamma^l_b l_b,
\end{aligned}
\right.
\end{align}
\end{subequations}
where 
\begin{align}
    & C^m_{ab} = \frac{\gamma}{M_0} \left[ \left( D({\bm q}) + J \right) M_0^2 + \left( D({\bm q}) + \frac{K_c}{2} \right) L_0^2  \right],
    \nonumber \\
    & C^m_{ba}=\gamma \left( D({\bm q}) + J + \frac{K_a}{2} \right) M_0,
    \nonumber \\
    & C^l_{ab} = \frac{\gamma}{M_0} \left[ D({\bm q}) M_0^2 + \left( D({\bm q}) + J + \frac{K_c}{2} \right) L_0^2 \right],
    \nonumber \\
    & C^l_{ba}=\gamma \left( D({\bm q}) + \frac{K_a}{2} \right) M_0,
    \nonumber
\end{align}
and the dumping parameters are $\Gamma^j_a = 2\alpha C^j_{ba}/M_0$ and $\Gamma^j_b = \alpha C_{ab}^j M_0/2$ with $j = m, l$. Here, we introduced the effective gyromagnetic ratio $\gamma = (\gamma_0 / n_s) / (1 + \alpha M_0^2)$ and defined the in-plane exchange part of the dispersion $D({\bm q})=A_a q_a^2 + A_{ab} q_a q_b + A_b q_b^2$. 

In accordance with previously accepted notations \cite{bae2022exciton,stetzuhn2025}, we recognize the ${\bm m}_\perp$-mode governed by Eq.~\eqref{eq:mdyn} as the out-of-phase magnon state (OP), for which the excitations of the layer magnetization vectors point in opposite directions, while the ${\bm l}_\perp$ mode is the in-phase (IP) magnon. The dispersion of these states reads: 
\begin{align}
    & \omega_j= \sqrt{C^j_{ab} C^j_{ba} - \frac{(\Gamma^j_a - \Gamma^j_b)^2}{4}} + i \frac{\Gamma^j_a + \Gamma^j_b}{2}.
    \label{om}
\end{align}

\begin{figure}[h!]
    \centering
    \includegraphics[width = \linewidth]{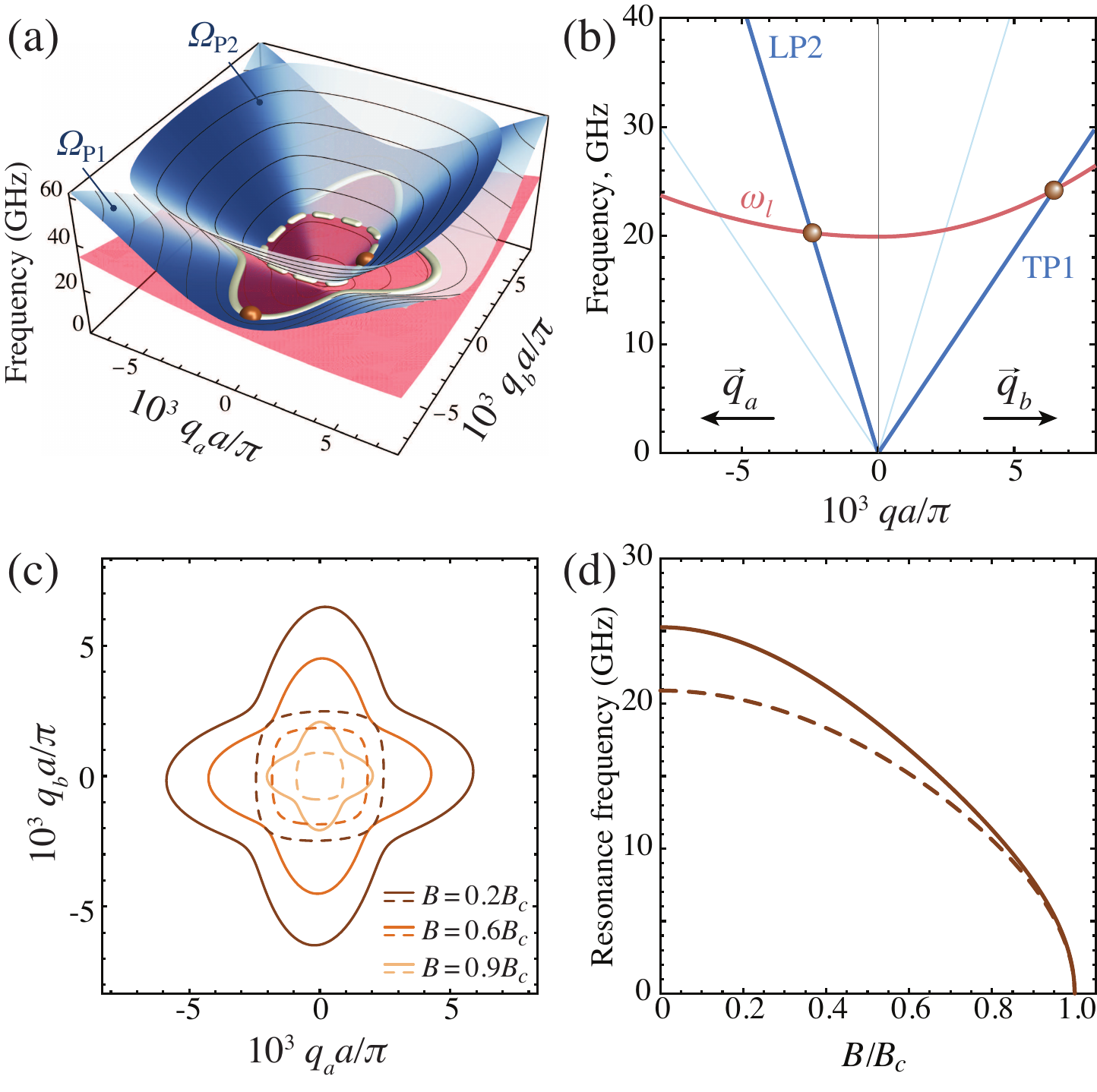}
    \caption{(a) Magnon and phonon dispersions in a CrSBr bilayer. The blue surfaces correspond to the lower $\Omega_{{\rm P}1}$ and upper $\Omega_{{\rm P}2}$ acoustic waves \eqref{phonon_disp}. The color saturation on the surface indicates the projection of the displacement vector $\bm{u}_0$ onto the $a$-axis. The red surface corresponds to the $l_\perp$ magnon mode at $B=0.2B_c$. The resonance conditions are shown with the gray solid and dashed lines. The points of the most efficient generation are shown by brown dots. (b) The  cross-section of the dispersion from panel (a) around the $\Gamma$-point. Magnons and phonons propagating in $b$-direction are shown on the right of the $\Gamma$-point, while the left side corresponds to the $a$-direction. The transverse TP1 and the longitudinal LP2 phonon modes with maximal $a$-component of the displacement magnitude $\bm{u}_0$ are shown with saturated blue color, while the other phonon states are shown with the pale shade. The resonances at the intersection of the magnon (red) and the phonon branches are indicated with brown dots. (c) The resonance cross-sections in the momentum space at various magnetic fields. The solid lines indicates the intersection with the lower phonon mode $\Omega_{{\rm P}1}$, while the dashed lines stand for the $\Omega_{{\rm P}2}$ resonance. (d) The dependence of resonant generation frequency corresponding to the dots in panels (a) and (b) on external magnetic field. The solid curve corresponds to the resonance with the TP1 phonon mode, while the gashed one stands for the LP2 resonance.}
\label{fig2}
\end{figure}

In the following, we focus on the ${\bm l}_\perp$-mode, whose dispersion is shown in red in Figs.~\ref{fig2}(a) and (b). The ${\bm m}_\perp$-mode remains decoupled from the elastic waves due to the specific symmetry of the magnetoelastic interaction in CrSBr. The interlayer exchange energy, $J {\bm n}_1 {\bm n}_2$, generates torques in the adjacent layer magnetizations that are equal in magnitude but opposite in direction: $\partial_t {\bm n}_{1,2} \propto \pm J \left[{\bm n}_{1} \times {\bm n}_{2}\right]$. Consequently, phonon-induced modulation of interlayer coupling does not excite net magnetization, leaving the ${\bm m}_\perp$-mode inactive.

The dispersion relation for phonons and their polarization can be obtained from the following equation for in-plane atomic displacements $\bm{u}$:
\begin{equation}
    \rho \frac{\partial^2 u_i}{\partial t^2} = C_{ijkl} \frac{\partial^2 u_l}{\partial x_j \partial x_k},
\label{phonon}
\end{equation}
where $\rho$ is the material density and $C_{ijkl}$ are the components of the elasticity tensor, symmetric in the permutation of the last two indices. 
The following non-zero values of the $\hat{C}$ tensor for an orthorhombic crystal (in Voigt notation) $C_{11} = 130$~GPa, $C_{22} = 129$~GPa, $C_{12} =15$~GPa, $C_{66} =26.5$~GPa were obtained by us using DFT calculations (see Supplemental Material \cite{SM}). Taking ${\bm u} = {\bm u}_0 e^{i({\bm q \bm r}-\omega t)}$, one obtains two phonon branches:
\begin{align}
    & \Omega_{{\rm P}1, {\rm P}2}^2 = \frac{1}{2\rho} \left[ 
    \vphantom{ 
    \sqrt{ \left((C_{11} - C_{66})q_a^2 - (C_{22} - C_{66})q_b^2\right)^2 - 4(C_{12} + C_{66})^2 q_a^2 q_b^2}
    }
    (C_{11} + C_{66})q_a^2 + (C_{22} + C_{66})q_b^2 \,\mp
    \nonumber \right. \\
    & \left. \sqrt{ \left((C_{11} - C_{66})q_a^2 - (C_{22} - C_{66})q_b^2\right)^2 + 4(C_{12} + C_{66})^2 q_a^2 q_b^2} \,\right],
\label{phonon_disp}
\end{align}
shown in Fig.~\ref{fig2}(a) with the blue color. The colormap encodes the projection of the displacement vector $\bm{u}_0$ onto the $a$-axis, which is the only strain component responsible for modulating the inter-layer exchange \cite{cenker2022reversible}. Therefore, magnons are excited most efficiently by the upper-branch longitudinal phonon (LP2) propagating along the $a$-axis with wave vector $\bm{q}_a$ or by the lower-branch transverse phonon (TP1) with $\bm{q} = \bm{q}_b$, see Figs.~\ref{fig2}(a) and (b). 

The efficient excitation requires matching of both energy and momentum between magnons and phonons. This matching occurs at the intersections of their dispersion surfaces, as illustrated in Fig.~\ref{fig2}(a) and in Fig.~\ref{fig2}(b) for the cross sections along the $\bm{q}_a$ (left, LP2 phonon) and $\bm{q}_b$ (right, TP1 phonon) directions.

The resulting resonance conditions form asymmetric contours in the wave vector plane, are shown in Fig.~\ref{fig2}(c), where solid and dashed lines correspond to resonances with the lower ($\Omega_{P1}$) and upper ($\Omega_{P2}$) branches, respectively. The shape and position of these contours are highly sensitive to the external magnetic field, enabling precise tuning of the energy and momentum of the generated magnons.
The magnetic field dependencies of the magnon frequencies at the TP1 and LP2 resonances are shown in Fig.~\ref{fig2}(d) with the solid and dashed lines, respectively.

Note that in addition to the resonances shown in Fig.~\ref{fig2}, additional intersections between the magnon and phonon dispersions occur in the sub-THz frequency range. We do not consider these resonances in this work, as generation of THz phonons is experimentally challenging.

To quantify the phonon-magnon coupling strength, we consider a weak periodic modulation of the interlayer coupling constant induced by the elastic wave: $J(t) = J_0 + W(t)$. The Neel vector then acquires an additional time-dependent torque $2 \,W(t)\left[{\bm n}_{1} \times {\bm n}_{2}\right] = W(t) [{\bm L} \times {\bm M}]$.
To leading order, ${\bm M}\approx M_0{\bm e}_c$ and ${\bm L}=L_0{\bm e}_b$, so this term acts as a driving force applied along the $a$-direction.
Eq.~\eqref{eq:ldyn} thus corresponds to one of a driven harmonic oscillator:
\begin{equation}
  \left\{
  \begin{aligned}
    \frac{dl_a}{dt} &= - C^l_{ab} l_y - \Gamma^l_a l_a + \gamma W({\bm r}, t) M_0 L_0,
    \\
    \frac{dl_b}{dt} &= C^l_{ba} l_a - \Gamma^l_b l_b.
\end{aligned}
\right.
\label{forced_oscillations}
\end{equation}
Note that the driving term exists only in the canted phase, where both $M_0$ and $L_0$ are nonzero, and vanishes in the collinear FM and AFM phases.

In what follows, we consider the driving acoustic wave $W({\bm r},t) = W_0 \cos({{\bm q r} - \Omega t})$ propagating with the wave vector $\bm{q}$ and frequency $\Omega\equiv \Omega(\bm{q}) $ given by Eq.~\eqref{phonon_disp}. Placing  $l_a = C_1(t){\bm A_1} e^{i \lambda_1 t} + C_2(t){\bm A_2} e^{i \lambda_2 t}$ into Eq.\eqref{forced_oscillations}, where $C_{1, 2}(t)$ are unknown functions, ${\bm A_{1, 2}}$ and $\lambda_{1, 2}$ are the eigenvectors and eigenvalues of the characteristic matrix 
\begin{equation}
\begin{pmatrix}
- \Gamma^l_a & - C^l_{ab} \\
C^l_{ba} & - \Gamma^l_b
\end{pmatrix}.
\nonumber
\end{equation}
one gets the amplitude $\ell$ of the $l_a$-component of the Neel vector:
\begin{equation}
    \ell = \frac{\gamma M_0 L_0 W_0 \sqrt{\Omega^2 + (\Gamma^{l}_b)^2}}{\sqrt{(\Omega^2 - \omega^2)^2 + 4 \Gamma^2 \Omega^2}},
\label{amp}
\end{equation}
where $\omega = | \omega_l | = \sqrt{(1+\alpha^2) C^l_{ab} C^l_{ba}}$ and $\Gamma = {\rm Im}(\omega_l) =  \left({\Gamma^l_a + \Gamma^l_b}\right)/{2}$.

\begin{figure}[h!]
    \centering
    \includegraphics[width = \linewidth]{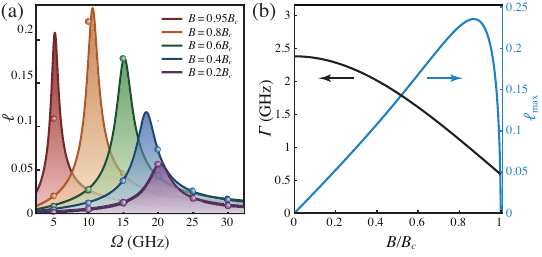}
    \caption{(a) The amplitude $\ell$ of the $l_a$-component of the IP spin wave excited by the LP2 phonons as a function of the driving frequency at various magnetic field strength. The driving strength amplitude is $0.1J_0$. The solid curves correspond to the predictions of the analytical model, Eq.~\eqref{amp}, while the results of the numerical simulation of the LLG dynamics are shown with the dots of the same color. The peak structure of the response curves indicates the resonant nature of the phonon-driven magnon generation in CrSBr bilayers. The resonance frequency decreases with increasing magnetic field in accordance with Fig.~\ref{fig2}(d). (b) Left axis -- The decay rate $\Gamma={\rm Im}(\omega_l) $ of the IP magnon mode at the resonance with LP2 phonons. This values characterizes a half-width of the resonance contours shown in panel (a). Right axis -- the peak oscillating amplitude $\ell_{\rm max}$ versus magnetic field strength. }
\label{fig:th+sim}
\end{figure}

\textit{Results}.
Below we investigate the dependence of $\ell$ on the acoustic wave frequency $\Omega$. To be specific, we consider the longitudinal phonons LP2 which propagate along the $a$-axis and assume that the amplitude of the $J$ modulation is 10\% of the equilibrium value, $W=0.1 J_0$. This is in accordance with recent observation of the switching in the interlayer magnetic order from the AFM to FM under the 0.7\% -1.2\% strain \cite{cenker2022reversible}. 
The resulting $\ell(\Omega)$-dependence predicted by Eq.~\eqref{amp} at $\omega = \Omega$ is shown in Fig.~\eqref{fig:th+sim} for different values of the magnetic field. It clearly demonstrates the resonant nature of the phonon-assisted magnon generation mechanism. 

To confirm the results of the linearized analytical model, we performed a numerical simulation of the LLG equations \cite{PhysRevB.108.174440}. The simulation was carried out on a one-dimensional uniform grid. We have considered the following magnetic field values: $0.2 B_c$, $0.4 B_c$, $0.6 B_c$, $0.8 B_c$, $0.95 B_c$. In each field, we varied the frequency of the exciting phonon from $5$ to $30$ GHz with the $5$ GHz step. The values of the $l_a$ amplitude established after a 2 ns long evolution, which is enough to reach a steady state, are shown in Fig.~\ref{fig:th+sim} with color dots. The analytical results demonstrate good agreement with simulations of the full LLG dynamics for the considered excitation amplitude.

The dependence of the peak generation amplitude and half-width of a resonant generation contour on the external magnetic field is shown in Fig.~\ref{fig:th+sim}(b) with blue and black curves, right and left axes, respectively. Note that the maximum of the blue curve, corresponding to the magnetic field optimal for magnon generation, is shifted in the region of larger magnetic fields with respect to those which maximize the coupling strength between phonons and magnons $B = \left({\sqrt{5}-1}\right)/{2} B_c \approx 0.62 B_c$, which is due to the decrease of the losses $\Gamma$ with the magnetic field, as shown by the black curve. 

The proposed mechanism can be experimentally tested using time-resolved pump-probe spectroscopy, which has been successfully employed to detect magnon dynamics in two-dimensional magnets \cite{bae2022exciton}. In the canted phase, the resonant modulation of interlayer exchange by an acoustic wave induces coherent oscillations of the Neel vector that modulate exciton energy via strong magneto-exciton coupling \cite{Fei2024}. The change in exciton energy $\Delta E_{ex}$ with a change in angle $\theta$ between ${\bf n}_1$ and ${\bf n}_2$ can be estimated as $\Delta E_{ex} = \Delta E^0_{ex} \cos^2{\theta / 2}$, where $E^0_{ex} = 20$ meV. In the magnetic field $0.8B_c$ and with a phonon frequency of 10 GHz, one obtains a modulation depth of $5.3$~meV. These effects should appear as periodic variations in reflectivity or magneto-optical Kerr effect (MOKE) signals at the acoustic frequency. The disappearance of the response in the collinear AFM and FM phases would provide a distinctive signature of the mechanism.

Note that we considered an out-of-plane orientation of the magnetic field. However, the proposed mechanism will also work in the case of a field applied along an intermediate magnetic axis. Such a realization would require smaller magnetic fields due to the lower anisotropy constant.

\textit{Conclusions}.
We proposed a new mechanism for the efficient resonant generation of magnons in CrSBr bilayers by acoustic waves, showing that the application of a magnetic field creates a canted magnetic phase that supports a forced magnetic oscillation mode. We demonstrated that the resonant frequency can be tuned in the range of 1--30 GHz. The creation of effective magnon sources can have a significant impact on the development of modern spintronics.

\textit{Acknowledgments.} The work was supported by the Russian Science Foundation Grant No. 25-72-20029. A.S. acknowledges the support of ’BASIS’ Foundation (Project No. 25-1-5-89-1).

\bibliography{bibliography}

\clearpage

\setcounter{equation}{0}
\setcounter{figure}{0}
\setcounter{table}{0}
\makeatletter
\renewcommand{\theequation}{S\arabic{equation}}
\renewcommand{\thefigure}{S\arabic{figure}}
\renewcommand{\thetable}{S\arabic{table}}

\section{Supplemental Material: Acoustically-driven magnons in CrSBr bilayers}

\subsection{Continuous model of interlayer exchange}

Using the intralayer exchange coupling constants $J_a = 4.27$~meV, $J_b= 3.76$~meV, $J_{ab}= 7.6$~meV from [22] we formulated the CrSBr monolayer energy lattice model in the following form:
\begin{align}
    & E = \frac{J_a}{2} \sum_{\bm R} {\bm S}_A ({\bm R}) {\bm S}_A ({\bm R + R}_a) + {\bm S}_A ({\bm R}) {\bm S}_A ({\bm R - R}_a) +
    \nonumber
    \\ 
    & {\bm S}_B ({\bm R}) {\bm S}_B ({\bm R + R}_a) + {\bm S}_B ({\bm R}) {\bm S}_B ({\bm R - R}_a) +
    \nonumber
    \\
    & \frac{J_b}{2} \sum_{\bm R} {\bm S}_A ({\bm R}) {\bm S}_A ({\bm R + R}_b) + {\bm S}_A ({\bm R}) {\bm S}_A ({\bm R - R}_b) +
    \nonumber
    \\
    & {\bm S}_B ({\bm R}) {\bm S}_B ({\bm R + R}_b) + {\bm S}_B ({\bm R}) {\bm S}_B ({\bm R - R}_b) +
    \nonumber
    \\
    & \frac{J_{ab}}{2} \sum_{\bm R} {\bm S}_A ({\bm R}) {\bm S}_B ({\bm R}) + {\bm S}_A ({\bm R}) {\bm S}_B ({\bm R - R}_b) +
    \nonumber
    \\
    & {\bm S}_A ({\bm R}) {\bm S}_B ({\bm R - R}_a) + {\bm S}_A ({\bm R}) {\bm S}_B ({\bm R} - {\bm R}_a - {\bm R}_b) +
    \nonumber
    \\
    & {\bm S}_B ({\bm R}) {\bm S}_A ({\bm R}) + {\bm S}_B ({\bm R}) {\bm S}_A ({\bm R + R}_b) +
    \nonumber
    \\
    & {\bm S}_B ({\bm R}) {\bm S}_A ({\bm R + R}_a) + {\bm S}_B ({\bm R}) {\bm S}_A ({\bm R} + {\bm R}_a + {\bm R}_b),
\end{align}
where ${\bm S}_A$ and ${\bm S}_B$ are the magnetizations of two different Cr atoms in a unit cell, ${\bm R}_a$ and ${\bm R}_b$ are the translation vectors in the crystallographic directions $a$ and $b$, ${\bm R}$ is the unit cell coordinate. 

Decomposing ${\bm S}({\bm R + \bm r})$ into a Taylor series, we found a connection between ${\bm S}({\bm R + \bm r})$ and ${\bm S}({\bm R})$:
\begin{align}
    & {\bm S}({\bm R + \bm r}) = {\bm S}({\bm R}) + \frac{\partial {\bm S}({\bm R})}{\partial R_a} r_a + \frac{\partial {\bm S}({\bm R})}{\partial R_b} r_b + 
    \nonumber
    \\
    & \frac{1}{2} \frac{\partial^2 {\bm S}({\bm R})}{\partial R_a^2} r^2_a + \frac{1}{2} \frac{\partial^2 {\bm S}({\bm R})}{\partial R_b^2} r^2_b + \frac{\partial^2 {\bm S}({\bm R})}{\partial R_a \partial R_b} r_a r_b.
\end{align}

In the ferromagnetic state ${\bm S}_A ({\bm R}) = {\bm S}_B ({\bm R})=S \, {\bm n} ({\bm R})$. Here we take into account that $S = S_A = S_B = 3/2$, $n = 1$. Using the resulting relationship and moving from summation to integration, we obtained the following expression for energy:

\begin{align}
    & E = \int d^2{\bm R} \left[ \frac{a}{b} \left( J_a + \frac{J_{ab}}{2} \right) \left( \frac{3}{2} \right)^2 \left( \frac{\partial {\bm n}}{\partial R_a} \right)^2 + \right.
    \nonumber
    \\ 
    & \left.\frac{b}{a} \left( J_b + \frac{J_{ab}}{2} \right) \left( \frac{3}{2} \right)^2 \left( \frac{\partial {\bm n}}{\partial R_b} \right)^2 + \right.
    \nonumber
    \\
    & \left.J_{ab} \left( \frac{3}{2} \right)^2 \frac{\partial {\bm n}}{\partial R_a} \frac{\partial {\bm n}}{\partial R_b} \right].
\end{align}

Then $A_a$, $A_b$ and $A_{ab}$ can be expressed in terms of $J_a$, $J_b$ and $J_{ab}$:

\begin{align}
    & A_a = \frac{a}{b} \left( J_a + \frac{J_{ab}}{2} \right) \left( \frac{3}{2} \right)^2,
    \\ 
    & A_b = \frac{b}{a} \left( J_b + \frac{J_{ab}}{2} \right) \left( \frac{3}{2} \right)^2,
    \\ 
    & A_{ab} = J_{ab} \left( \frac{3}{2} \right)^2.
\end{align}

\subsection{Full system of LLG equations}

The energy of the system $E_0$ after the transition to the variables ${ \bm M}$ and ${ \bm L}$ has the following form:
\begin{align}
    & E_0 =  \int {\rm d}^2 {\bm r} \left( \frac{1}{2} (\hat{D}[{\bf M}] + \hat{D}[{\bf L}]) + \frac{K_c}{4} (M^2_{c} + L^2_{c}) + \right.
    \nonumber \\
    & \left.\frac{K_a}{4} (M^2_{a} + L^2_{a}) + \frac{J}{2}{\bm M^2} - BM_c  \right).
\end{align}

Then the system of LLG equations can be written as follows: 
\begin{align}
\begin{cases}
    \frac{\partial {\bm M}}{\partial t} = \gamma \left[ {\bm M} \times \frac{ \delta E_0}{\delta {\bm M}} + {\bm L} \times \frac{ \delta E_0}{\delta {\bm L}}  \right] \\
    + \gamma \frac{\alpha}{2} \left[ {\bm M} \times \left[ {\bm M} \times\frac{ \delta E_0}{\delta {\bm M}} \right] + {\bm L} \times \left[ {\bm L} \times \frac{ \delta E_0}{\delta {\bm L}}  \right] \right. \\
    + \left.{\bm M} \times \left[ {\bm L} \times\frac{ \delta E_0}{\delta {\bm L}} \right] + {\bm L} \times \left[ {\bm M} \times\frac{ \delta E_0}{\delta {\bm L}} \right] \right], \\
    \noalign{\vspace{+0.5ex}}
    \frac{\partial {\bm L}}{\partial t} = \gamma \left[ {\bm M} \times \frac{ \delta E_0}{\delta {\bm L}} + {\bm L} \times \frac{ \delta E_0}{\delta {\bm M}}  \right] \\
    + \gamma \frac{\alpha}{2} \left[ {\bm M} \times \left[ {\bm M} \times\frac{ \delta E_0}{\delta {\bm L}} \right] + {\bm L} \times \left[ {\bm L} \times \frac{ \delta E_0}{\delta {\bm M}}  \right] \right. \\
    + \left.{\bm M} \times \left[ {\bm L} \times\frac{ \delta E_0}{\delta {\bm M}} \right] + {\bm L} \times \left[ {\bm M} \times \frac{ \delta E_0}{\delta {\bm M}}  \right] \right].
\end{cases}
\end{align}
The dynamics of ${ \bf m}$ and ${ \bf l}$ is described by the system
\begin{align}
\begin{cases}
    \frac{dm_a}{dt} = - \gamma \left[ \left( D({\bf q}) + J \right) M_0 m_b - \left( D({\bf q}) + \frac{K_c}{2} \right) L_0 l_c \right] - 
    \\
    \gamma \frac{\alpha}{2} \left( D({\bf q}) + J + \frac{K_a}{2} \right) (M_0^2 + L_0^2) m_a,
    \\
    \frac{dm_b}{dt} = \gamma \left[ \left( D({\bf q}) + J + \frac{K_a}{2} \right) M_0 m_a \right] -
    \\
    \gamma \frac{\alpha}{2} \left( D({\bf q}) + J \right) M_0^2 m_b + \gamma \frac{\alpha}{2} \left( D({\bf q}) + \frac{K_c}{2} \right) M_0 L_0 l_c,
    \\
    \frac{dm_c}{dt} = - \gamma \left[ \left( D({\bf q}) + \frac{K_a}{2} \right) L_0 l_a \right] -
    \\
   \gamma \frac{\alpha}{2} \left( D({\bf q}) + J + \frac{K_c}{2} \right) L_0^2 m_c + \gamma \frac{\alpha}{2} D({\bf q}) M_0 L_0 l_b,
    \\
    \frac{dl_a}{dt} = - \gamma \left[ D({\bf q}) M_0 l_b - \left( D({\bf q}) + J + \frac{K_c}{2} \right) L_0 m_c \right] -
    \\
    \gamma \frac{\alpha}{2} \left( D({\bf q}) + \frac{K_a}{2} \right) (M_0^2 + L_0^2) l_a,
    \\
    \frac{dl_b}{dt} = \gamma \left[ \left(D({\bf q}) + \frac{K_a}{2} \right) M_0 l_a \right] -
    \\
    \gamma \frac{\alpha}{2} D({\bf q}) M_0^2 l_b + \gamma \frac{\alpha}{2} \left( D({\bf q}) + J + \frac{K_c}{2} \right) M_0 L_0 m_c,
    \\
    \frac{dl_c}{dt} = - \gamma \left[ \left( D({\bf q}) + J + \frac{K_a}{2} \right) L_0 m_a \right] -
    \\
    \gamma \frac{\alpha}{2} \left( D({\bf q}) + \frac{K_c}{2} \right) L_0^2 l_c + \gamma \frac{\alpha}{2} \left( D({\bf q}) + J \right) M_0 L_0 m_b.
\end{cases}
\end{align}

Using the normalization conditions, we derived Eq. (3a) and Eq. (3b) from this system.

\subsection{DFT calculations of elastic constants}

First, the positions of the atoms inside the CrSBr monolayer unit cell were relaxed using the BFGS (Broyden—Fletcher—Goldfarb—Shanno) method. The calculations were performed using the GPAW package [29-31]. We used the optB88-vdW functional from the libvdwxc library [32] and a 8x8x1 k-point mesh in the first Brillouin zone. We set the plane-wave energy cutoff equal to 900 eV and added a vacuum layer 15 angstrom in the c direction to avoid interfacial interactions.

Elastic constants were calculated using the  finite deformation method. The calculations were performed using the elastic package [33-35] and the PBE exchange-correlation functional [36]. 

\end{document}